\documentclass[twocolumn,prb,showpacs,floatfix]{revtex4}

\usepackage{graphicx}
\usepackage{dcolumn}
\usepackage{float}
\usepackage{amsmath}

\newcommand{\comment}[1]{}

\begin{document}


\title{Optical conductivity of bilayer graphene with and without
an asymmetry gap}

\author{E.J. Nicol}
\affiliation{Department of Physics, University of Guelph,
Guelph, Ontario, N1G 2W1, Canada}
\email{nicol@physics.uoguelph.ca}
\author{J.P. Carbotte}
\affiliation{Department of Physics and Astronomy, McMaster University,
Hamilton, Ontario, L8S 4M1, Canada}

\date{\today}

\begin{abstract}
{When a bilayer of graphene is placed in a suitably configured field
effect device, an asymmetry gap can be generated and the carrier
concentration made different in each layer. This provides a tunable
semiconducting gap, and the valence and the conductance band
no longer meet at the two Dirac points of the graphene Brillouin
zone. We calculate the optical conductivity of such a semiconductor
with particular emphasis on the optical spectral weight redistribution brought
about by changes in gap and chemical potential due to charging. 
We derive
an algebraic formula for arbitrary value of the chemical potential
for the case of the bilayer conductivity without
a gap. 
}
\end{abstract}


\maketitle

\section{Introduction}

Graphene is known to exhibit special properties related to the Dirac
nature of its quasiparticle dynamics. As an example, a half-integer
quantum Hall effect is observed\cite{Novoselov,Zhang} as was 
predicted\cite{Gusynin,Neto}. Bilayer graphene also possesses remarkable
properties. When placed in a suitably configured field effect device,
a tunable semiconducting gap can be generated with the result
that the valence and the conduction band no longer meet at the
two Dirac points in the graphene Brillouin 
zone\cite{Ohta,Castro,
McCann,McCann2,Nilsson}. For a review of other remarkable 
properties of such systems as well as a discussion of possible technological
applications, the reader is referred to Ref.~\cite{Geim,NetoRMP}.
The optical conductivity of a few-layer epitaxial 
graphite\cite{Sadowski1,Sadowski2} and oriented 
pyrolytic graphite\cite{Li,Kuzmenko}
in finite external magnetic field has been reported recently, as well
as for graphene\cite{Jiang}. There have also been theoretical 
studies\cite{Nilsson,Zheng,Abergel,Falkovsky,Gusynin2} of the 
conductivity, including discussions of optical sum rules\cite{Gusynin3,Benfatto}
which continue to provide useful information\cite{Carbotte} on the
electron dynamics. In Ref.~\cite{Benfatto} it was found that the opening of
the asymmetry gap in bilayer graphene leads to very small changes in 
the overall optical sum. Here we consider the optical conductivity
of such a system with particular emphasis on the optical spectral weight
redistribution brought about by changes in the chemical potential, due to
charging, and to the opening of a semiconducting gap. In the configuration
envisioned here, donor atoms are seeded on the upper face of a
bilayer which is also placed in a field effect device so that the
carrier imbalance in each layer is different as is the electrostatic
potential.

The structure of our paper is organized as follows. In section II,
we outline the theoretical derivation of the optical conductivity
starting from the simplest nearest-neighbor tight-binding Hamiltonian
which includes terms associated with the biased bilayer configuration.
In section III, we present results for the case without the anisotropy gap,
pertinent to the unbiased graphene bilayer. Here, we have been able to
derive a surprisingly simple analytical formula for the frequency-dependent
conductivity for arbitrary choice of chemical potential and we have tested
it against the full numerical solution and find good agreement, making this
a very useful formula for experimentalists. In pure monolayer graphene,
the conductivity is flat for the half-filled case of $\mu=0$ and has a
universal value of $\pi e^2/2h$. With finite $\mu$, the spectral weight
below a frequency $\Omega=2\mu$ is transferred to a Drude metallic response at
zero frequency.\cite{Gusynin4} We find in the case of the pure bilayer,
for the half-filled $\mu=0$ case, the conductivity retains a universal
value at $\omega=0$ and at high frequency, which is now twice the monolayer
case, i.e. $\pi e^2/h$. Structure in the conductivity is found at $\Omega=\gamma$
and $2\gamma$, with the structure at $\gamma$ being particularly pronounced
due to  a doubly degeneratre set of interband transitions. With finite
$\mu$, the doubly degnerate set of transitions is split into two separate
features and a new strong absorption at $\gamma$ is seen due to the nesting
of two energy bands. At low energy, spectral weight is once again removed
below $2\mu$ and a Drude peak occurs. In section IV, we consider the case 
of the bilayer with an asymmetry gap. The main results in this case
is the appearance of a semiconducting gap in the conductivity for $\mu=0$
and a shift in the structure of the doubly degenerate  interband transitions
to higher energy. For finite $\mu$, the metallic behavior is restored and a Drude
peak occurs accompanied by loss of spectral weight below $2\mu$ and the structure 
due to the doubly degenerate transitions is split, as before. The nesting
feature at $\gamma$ in the unbiased bilayer is now shifted to a new value
and is broadened due to imperfect nesting of the energy bands. Finally,
we end our analysis with a discussion in Section V.

\section{Theoretical Background}

In order to derive and discuss the optical conductivity of biased
bilayer
graphene, we need to examine the form of the band structure
and provide an expression for the electronic Green's function.
To this end, we begin with the Hamiltonian for the system under
consideration and follow the notation given by McCann\cite{McCann}
and Benfatto et
al.\cite{Benfatto}
for continuity. The single spin Hamiltonian is given as:
\begin{widetext}
\begin{eqnarray}
H&=&-t\sum_{\bf n,\boldsymbol{\delta}} (a_{1{\bf n+\boldsymbol{\delta}}}^\dagger b_{1\bf n}
+h.c.)
-t\sum_{\bf n,\boldsymbol{\delta}^\prime} (a_{2{\bf
    n}}^\dagger b_{2\bf n+\boldsymbol{\delta}^\prime} +h.c.)
+\gamma\sum_{\bf n}(a^\dagger_{2\bf n}b_{1\bf n}+h.c.)
\nonumber\\
&-&\frac{1}{2}\Delta\sum_{\bf n}(a^\dagger_{1{\bf n}+\boldsymbol{\delta}_1}
a_{1{\bf n}+\boldsymbol{\delta}_1}
+b^\dagger_{1\bf n}b_{1\bf n})
+\frac{1}{2}\Delta\sum_{\bf n}(a^\dagger_{2\bf n}a_{2\bf
  n}
+b^\dagger_{2{\bf n}+\boldsymbol{\delta}_1^\prime}b_{2{\bf n}+\boldsymbol{\delta}_1^\prime}).
\label{eq:H}
\end{eqnarray}
\end{widetext}
The first two terms are the nearest neighbor hopping terms for
electrons to move in each of the graphene planes, separately.
The two planes are indexed by 1 and 2, with a single graphene sheet having
two inequivalent atoms labelled $A$ and $B$, as the arrangement of carbon 
atoms on the two-dimensional honeycomb lattice provides for two atoms
per unit cell. The operator $b_{1\bf n}$ annihilates an electron
on the $B$-atom site with site label ${\bf n}$ in the graphene sheet
with
label 1 and $a^\dagger_{1\bf n+\boldsymbol{\delta}}$ creates an electron on the
neighboring $A$-atom site in the same sheet positioned at ${\bf
  n+\boldsymbol{\delta}}$,
where $\boldsymbol{\delta}$ can be one of three vectors which point to
the three possible nearest neighbors. These vectors are enumerated
as $\boldsymbol{\delta}_1=-({\bf a}_1+{\bf a}_2)/3$, 
$\boldsymbol{\delta}_2=(2{\bf a}_1-{\bf a}_2)/3$,
and $\boldsymbol{\delta}_3=-({\bf a}_1-2{\bf a}_2)/3$,  
where ${\bf a}_1=(a\sqrt{3}/2,a/2)$ and ${\bf a}_2=(a\sqrt{3}/2,-a/2)$ are
the unit vectors of the triangular sublattice for the $A$ or $B$
atoms, and
$a=|{\bf a}_1|=|{\bf a_2}|=\sqrt{3}a_{CC}$ 
with $a_{CC}$ the distance between two nearest carbon atoms.
Now in layering graphene sheets,
there are several choices for stacking. The one
under consideration here and in experiment, is that of the Bernal-type
stacking where if the atoms are labelled $A1$ and $B1$ in sheet 1
and $A2$ and $B2$ in sheet 2, then the $A2$ atoms
are
stacked directly over the $B1$ atoms, but the $B2$ atoms are 
stacked over the centers of the $A1$-$B1$ carbon rings.
Pictures of this structure can be found in several references\cite{McCann2,NetoRMP}
and so we do not reproduce this here. Note, however, that if we
use the index ${\bf n}$ to reference the $B1$ atoms, then transferring this
index to the $A2$ atoms directly above gives rise to indexing
the nearest neighbor vectors in the second sheet relative to the
$A$ atoms and hence the vectors
are orientated differently and labelled as
${\boldsymbol{\delta}_i^\prime}=-\boldsymbol{\delta}_i$.
This stacking means that in terms of 
nearest neighbors associated with interlayer coupling, the $A2$
atom is the nearest neighbour of the $B1$ atom through a direct
vertical
bond. Thus, in the Hamiltonian of Eq.~(\ref{eq:H}), the third term
shows the hopping term for an electron on the $B1$ site to hop
to the $A2$ site in plane 2. The hopping parameter is given as 
$\gamma$ and is typically about 0.4 eV.
There is also a possibility to hop from $B1$ to $B2$
or $A1$ to $A2$ but these hopping energies are very
small (see Ref.~\cite{McCann2} and references therein). 
Hopping from $A1$ to $B2$ 
in the Hamiltonian is another possibility and its energy is
larger at $\sim 0.315$ eV,\cite{McCann2} 
however, we do not display it in the Hamiltonian
as we drop this term to focus on the physics and results which
come from the main interlayer coupling,
the $\gamma$ term, which is essential to the discussion of
the bilayer configuration. Furthermore, we are considering
a biased bilayer structure which can give rise to the very
novel physics of being able to tune the bilayer from
metallic to semiconducting behavior, consequently the biasing
is indicated in Eq.~(\ref{eq:H}) as a lowering of the energy
on graphene plane 1 by an amount $\Delta/2$ and the raising
of the energy on plane 2 by the same amount. This results in
the last two terms of Eq.~(\ref{eq:H}). The quantity $\Delta$
is referred to as the asymmetry gap.
The Hamiltonian transforms in the standard way\cite{Mahan}
to $k$-space and can be written as
a matrix:
\begin{equation}
\hat H=\left(\begin{array}{cccc}
-\frac{\Delta}{2} & 0                  & 0                  &\phi^*({\bf k})\\
0                  & \frac{\Delta}{2} & \phi({\bf k})     & 0\\
0                  & \phi^*({\bf k})   & \frac{\Delta}{2} & \gamma\\
\phi({\bf k})     & 0                  & \gamma          & -\frac{\Delta}{2}
\end{array}\right) ,
\label{eq:Hmatrix}
\end{equation}
where $\phi({\bf k})=-t\sum_{\bf \boldsymbol{\delta}}e^{i{\bf k\cdot\boldsymbol{\delta}}}=
-t\sum_{\bf \boldsymbol{\delta}^\prime}e^{-i{\bf k\cdot\boldsymbol{\delta}^\prime}}$ and we have
followed the notation of McCann\cite{McCann} by using an eigenvector
$\Psi=(a_{1\bf k},b_{2\bf
  k},a_{2\bf k},b_{1\bf k})$.
The energy eigenvalues of this matrix define the band structure. As
the bilayer has four atoms per unit cell (doubling that of single
layer graphene), there are now four bands and these are given by:
\begin{eqnarray}
\epsilon^2_\alpha(\bf k)&=&\frac{\gamma^2}{2}
+\frac{\Delta^2}{4}+|\phi({\bf
    k})|^2+(-1)^\alpha\Gamma ,\nonumber\\
\Gamma&=& \sqrt{\frac{\gamma^4}{4}+|\phi({\bf
    k})|^2(\gamma^2+\Delta^2)},
\label{eq:band}
\end{eqnarray}
where $\alpha=1$, 2.
Of these four bands, the lower energy ones $\pm\epsilon_1({\bf k})$
are essentially the original graphene bands with  low energy
modification and the $\pm\epsilon_2({\bf k})$ are higher energy
bands reflecting the dimerized bond between $B1$ and $A2$,
which has an energy scale of $\gamma$. As the main low energy 
physics occurs at the two inequivalent $K$ and $K^\prime$
points of the graphene Brillouin zone, the function
$\phi({\bf k})$ can be expanded around the $K$ point in the
continuum 
approximation (i.e., the limit of
small lattice constant $a$) to be $|\phi({\bf k})|\approx\hbar v_F{\bf k}$,
where $v_F=\sqrt{3}ta/(2\hbar)$. With this approximation, if $\Delta=0$ and
$\gamma=0$, we would recover the famous graphene band structure
where $\epsilon=\pm \hbar v_F{\bf k}$ is the form of the dispersion
around the so-called Dirac points, which are two-fold degenerate
for the uncoupled bilayer. However, if $\Delta=0$ and $\gamma\ne 0$,
the band structure around these points is modified to be quadratic
in $k$ (although still linear at larger $k$) and the degeneracy is
lifted such that the dimerized bands are shifted by $\gamma$,
as can been seen later on in our first figure. The presence of the
bias energy $\Delta$, produces an energy gap in the band structure and
a 
``mexican hat'' structure occurs in the lower energy band $\epsilon_1$
with a minimum at $E_{g1}=\gamma\Delta/(2\sqrt{\gamma^2+\Delta^2})$
at $|\phi({\bf k})|= \phi(k_0)=\hbar v_Fk_0$,
where $k_0=(\Delta/2)\sqrt{(\Delta^2+2\gamma^2)/(\Delta^2+\gamma^2)}$ 
and a ``hat'' maximum at $E_{01}=\Delta/2$
for $k=0$. This will be discussed further in Section~IV.
This unusual band structure and the presence of the energy 
scales of $\gamma$ and $\Delta$,
gives rise to very rich structure
in the frequency-dependent conductivity, as we will see.

With this Hamiltonian, it is straightforward to obtain the 
Green's function $\hat G$ through $\hat G^{-1}=z\hat I-\hat H$ or
\begin{equation}
\hat G^{-1}(z)=\left(\begin{array}{cccc}
z+\frac{\Delta}{2} & 0                  & 0                  &-\phi^*({\bf k})\\
0                  & z-\frac{\Delta}{2} & -\phi({\bf k})     & 0\\
0                  & -\phi^*({\bf k})   & z-\frac{\Delta}{2} & -\gamma\\
-\phi({\bf k})     & 0                  & -\gamma          & z+\frac{\Delta}{2}
\end{array}\right) ,
\label{eq:Ginv}
\end{equation}
where $z=i\omega_n$, with $\omega_n=\pi T(2n+1)$ the fermionic Matsubara
frequency for $n=0,\pm 1,\pm 2,\dots$ and $T$, the
temperature. For our calculation
of the optical conductivity, only certain elements of the Green's
function enter our final expression and they are: $G_{11}$, $G_{22}$,
$G_{33}$, $G_{44}$, $G_{13}$, $G_{24}$. As $G_{22}(\Delta)=G_{11}(-\Delta)$, 
$G_{33}(\Delta)=G_{44}(-\Delta)$, and $G_{24}(\Delta,\phi)=G_{13}(-\Delta,\phi^*)$,
it is sufficient to show only three elements explicitly:
\begin{eqnarray} 
G_{11}&=&\frac{(\Delta-2z)(\Delta^2+4\gamma^2-4z^2)
-4|\phi({\bf
  k})|^2(\Delta+2z)}{8(z^2-\epsilon_1^2)(z^2-\epsilon_2^2)},\\
G_{44}&=&\frac{(\Delta+2z)[(\Delta-2z)^2
-4|\phi({\bf k})|^2]}{8(z^2-\epsilon_1^2)(z^2-\epsilon_2^2)},\\
G_{13}&=&\frac{(2z-\Delta)\gamma\phi({\bf k})}
{2(z^2-\epsilon_1^2)(z^2-\epsilon_2^2)}.
\label{eq:green}
\end{eqnarray}

The finite frequency conductivity is calculated through the standard
procedure of using the Kubo formula\cite{Mahan}. The real part of
the conductivity is written in terms of the retarded current-current
correlation function
$\Pi_{\alpha\beta}(\Omega+i0^+)$ as
\begin{equation}
\sigma_{\alpha\beta}(\Omega)= \frac{{\rm Im}
  \Pi_{\alpha\beta}(\Omega+i0^+)}{\Omega} ,
\label{eq:kubo1}
\end{equation}
where $\alpha$ and $\beta$ indicate the spatial components
(here we will be interested in the longitudinal conductivity 
parallel to the graphene sheets, $\sigma_{xx}$). The 
retarded current-current correlation
function is also referred to as the polarization
 function which we calculate
in Matsubara formalism as outlined by Mahan\cite{Mahan}
and given as
\begin{equation}
\Pi_{\alpha\beta}(i\nu_m)=-\int_{0}^{1/T} d\tau e^{i\nu_m\tau}
<T_\tau J_\alpha(\tau)J_\beta(0)>,
\label{eq:Pi}
\end{equation}
where $\tau$ is imaginary time, $T_\tau$ is  the time ordering
operator  and $\nu_m$ is the bosonic Matsubara frequency $2\pi mT$
for $m=0,\pm 1,\pm 2,\dots$ and temperature $T$. In Eq.~(\ref{eq:kubo1})
$\Pi_{\alpha\beta}(\Omega+i0^+)$ is the analytic continuation
of Eq.~(\ref{eq:Pi}) to the real axis via $i\nu_m\to\Omega+i0^+$.
The current operator is the sum over sites of the site-specific
paramagnetic current operator\cite{Gusynin3}:
\begin{equation}
J_\alpha(\tau)=\sum_{\bf n}j^P_\alpha(\tau,{\bf n}).
\end{equation}
In order to evaluate this, we require the particular current operator
corresponding to our Hamiltonian. This is found via a Peierls
substitution on Eq.~(\ref{eq:H}), where the operators associated
with hopping to a new site are modified as
$a_{1{\bf n+\boldsymbol{\delta}}}^\dagger b_{1\bf n}
\to a_{1{\bf n+\boldsymbol{\delta}}}^\dagger
\exp(-\frac{ie}{\hbar}\int_{\bf n}^{\bf n+\boldsymbol{\delta}}{\bf
  A}\cdot
{\bf r}
) b_{1\bf n}$. (Note that we have taken the velocity of light $c=1$.)
The new Hamiltonian with
the vector potential ${\bf A}$ is
then expanded for small ${\bf A}$ to first order and the paramagnetic
current operator is given by 
$j^P_\alpha({\bf n})=-\partial H/\partial(A_\alpha({\bf n}))$.
For our Hamiltonian and the case of considering only
currents in the graphene sheets, the paramagnetic current is given as
\begin{eqnarray}
j_\alpha^P({\bf n})&=&-\frac{ite}{\hbar }
\sum_{\boldsymbol{\delta}} (\boldsymbol{\delta})_\alpha
(a_{1{\bf n+\boldsymbol{\delta}}}^\dagger b_{1\bf n}
-h.c.)\nonumber\\
&&+\frac{ite}{\hbar }\sum_{\boldsymbol{\delta}^\prime} 
(\boldsymbol{\delta}^\prime)_\alpha (a_{2{\bf
    n}}^\dagger b_{2\bf n+\boldsymbol{\delta}^\prime} -h.c.).
\label{eq:jpara}
\end{eqnarray}
Note that the Hermitian conjugate piece has a minus sign representing
a depletion of the current for hopping in the reverse direction. 
Fourier transforming to $k$-space and summing over ${\bf n}$, 
we can write the total paramagnetic current operator 
$J_\alpha({\bf k})=-e\sum_{\bf k} \Psi^\dagger \hat v_\alpha \Psi$,
where
\begin{equation}
\hat v_\alpha =\left(\begin{array}{cccc}
0  & 0  & 0  &v_{\alpha\bf k}^*\\
0  & 0  & v_{\alpha\bf k}& 0 \\
0  & v_{\alpha\bf k}^*  & 0  &0\\
v_{\alpha\bf k}  & 0  & 0  & 0
\end{array}\right) ,
\label{eq:vmatrix}
\end{equation}
with $v_{\alpha\bf k}=-(it/\hbar)\sum_{\boldsymbol{\delta}}
(\boldsymbol{\delta})_\alpha e^{i{\bf k}\cdot{\boldsymbol{\delta}}}$.
The
structure of the equation embodies the form of $j=-nev$ where the
velocity $v$ is proportional to 
the gradient of the energy. One can see that the
factors
of $it$ and the sum over components of $\boldsymbol{\delta}$ 
weighted by $e^{i{\bf k}\cdot{\boldsymbol{\delta}}}$ are
simply a result of taking the gradient of $\phi({\bf k})$ which is
the bare energy in the absence of $\Delta$ and $\gamma$, i. e.,
$v_{\bf k}=\nabla \phi({\bf k})/\hbar$.
Consequently, the polarization function can then be written in the
usual bubble approximation as
\begin{widetext}
\begin{equation}
\Pi_{\alpha\beta}(i\nu_m)
=e^2T\sum_{i\omega_n}
\int \frac{d^2k}{(2\pi)^2}{\rm Tr}[\hat v_\alpha
 \hat G(i\omega_n+i\nu_m,{\bf k})\hat v_\beta \hat G(i\omega_n,{\bf k})]
\end{equation}
\end{widetext}
or, favoring the spectral function representation of the Green's
function where 
\begin{equation}
G_{ij}(z)=\int^{\infty}_{-\infty}
\frac{d\omega}{2\pi}\frac{A_{ij}(\omega^\prime)}{z-\omega^\prime} ,
\end{equation}
we can write the real part of the conductivity as
\begin{widetext}
\begin{equation}
\sigma_{\alpha\beta}(\Omega)
=\frac{e^2}{2\Omega}\int^{\infty}_{-\infty}\frac{d\omega}{2\pi}
[f(\omega-\mu)-f(\omega+\Omega-\mu)]
\int \frac{d^2k}{(2\pi)^2}{\rm Tr}[\hat v_\alpha
  \hat A(\omega+\Omega,{\bf k})\hat v_\beta \hat A(\omega,{\bf k})],
\label{eq:cond1}
\end{equation}
\end{widetext}
where $f(x)=1/[\exp(x/T)+1]$ is the Fermi function and
$\mu$ is the chemical potential. In presenting
our results, we evaluate this equation at T=0. 
Taking the trace and dropping those terms which vanish
in the averaging over momentum, the structure of the
equation to be evaluated reduces to knowing the three spectral
functions associated with the three Green's functions mentioned
earlier. The longitudinal conductivity
$\sigma_{xx}(\omega)\equiv\sigma(\Omega)$ becomes
\begin{widetext}
\begin{eqnarray}
\sigma(\Omega)
&=&\frac{N_fe^2}{2\Omega}\int^{\infty}_{-\infty}\frac{d\omega}{2\pi}
[f(\omega-\mu)-f(\omega+\Omega-\mu)]
\int \frac{d^2k}{(2\pi)^2}|v_{\bf k}|^2\bigl\{
A_{11}(\omega,\Delta)A_{44}(\omega+\Omega,\Delta)
+A_{44}(\omega,\Delta)A_{11}(\omega+\Omega,\Delta)\nonumber\\
&&+A_{11}(\omega,-\Delta)A_{44}(\omega+\Omega,-\Delta)
+A_{44}(\omega,-\Delta)A_{11}(\omega+\Omega,-\Delta)\nonumber\\
&&+2[A_{13}(\omega,\Delta)A^*_{13}(\omega+\Omega,-\Delta)+A^*_{13}(\omega,-\Delta)A_{13}(\omega+\Omega,\Delta)]\bigr\}.
\label{eq:cond}
\end{eqnarray}
\end{widetext}
We evaluate this in the continuum approximation around the
$K$ point of the graphene Brillouin zone, where $|v_{\bf k}|^2=v_F^2$
and $|\phi({\bf k})|=\hbar v_F k$ and the integral over $k$ has a large
upper cutoff typical of the large bandwidth.
Thus, we introduce a factor $N_f=4$ which comes from a sum over spin 
(not included explicitly up till now) and a
sum over the two inequivalent $K$ points ($K$
and $K^\prime$) in the graphene Brillouin zone. 
Previous to Eq.~(\ref{eq:cond}), the integral over $d^2k$ was to be
taken over the Brillouin zone, but in Eq.~(\ref{eq:cond}) it is to
be interpretted as over a single $K$ point which would be a cone for
the decoupled bilayer graphene.
If we take the limit
of this expression (Eq~(\ref{eq:cond})) for $\gamma=\Delta=\mu=0$, 
we find a constant as a function
of frequency given as $\sigma_0=e^2/2\hbar$, which is
twice the result for single layer graphene\cite{Gusynin4}.
We will refer to this
as the conductivity of the uncoupled graphene bilayer. 
Finally, for illustration, we write an example of the spectral functions here:
\begin{equation}
A_{13}(\omega,\Delta)=\sum_{\alpha=1,2}
[a_{13}(\alpha,\Delta)\delta(\omega-\epsilon_\alpha)
+a_{13}(\alpha,-\Delta)\delta(\omega+\epsilon_\alpha)],
\end{equation}
where
\begin{equation}
a_{13}(\alpha,\Delta)=(-1)^\alpha\frac{\pi}{2}\frac{(2\epsilon_\alpha-\Delta)\gamma
\phi({\bf k})}{\epsilon_\alpha(\epsilon_2^2-\epsilon_1^2)}.
\end{equation}
The other spectral functions follow from the Green's functions in a
similar
manner.
For numerical
work, we write the delta functions in the spectral functions 
as Lorentzians with a broadening $\eta$, i.e.,
$\delta(x)=(\eta/\pi)/[x^2+\eta^2]$. In the optical
conductivity, this manifests itself as an effective transport
scattering
rate of $1/\tau_{imp}=2\eta$ due to the convolution of two delta
functions
in the conductivity formula.

\section{Results without anisotropy gap}

At zero temperature, for the case of $\Delta=0$, it is possible to
derive a closed algebraic formula for the bilayer conductivity.
It has the form:
\begin{eqnarray}
\frac{\sigma(\Omega)}{\sigma_0}&=&
\biggl[\frac{\Omega+2\gamma}{2(\Omega+\gamma)}
+\frac{\Omega-2\gamma}{2(\Omega-\gamma)}\Theta(\Omega-2\gamma)
\biggr]\Theta(\Omega-2\mu)\nonumber\\
&+&
\frac{\gamma^2}{2\Omega^2}[\Theta(\Omega-2\mu-\gamma)
+\Theta(\Omega-2\mu+\gamma)]\Theta(\Omega-\gamma)
\nonumber\\
&+& a(\mu)\delta(\Omega)+b(\mu)\delta(\Omega-\gamma)
\label{eq:falkomu}
\end{eqnarray}
with
\begin{equation}
a(\mu)=\frac{4\mu(\mu+\gamma)}{2\mu+\gamma}
+\frac{4\mu(\mu-\gamma)}{2\mu-\gamma}\Theta(\mu-\gamma)
\label{eq:amu}
\end{equation}
and
\begin{equation}
b(\mu)=\frac{\gamma}{2}\biggl[\ln\frac{2\mu+\gamma}{\gamma}
-\ln\frac{2\mu-\gamma}{\gamma}\Theta(\mu-\gamma)\biggr]
\label{eq:bmu}
\end{equation}
where $\sigma_{0}=e^2/2\hbar$ which is twice the conductivity of a single graphene
sheet. This expression correctly reduces to the form given by Abergel
and Fal'ko\cite{Abergel} in the limit of $\mu=0$. In the top frame of
Fig.~\ref{fig1}, we show results for $\sigma(\Omega)/\sigma_0$ as
a function of $\Omega/\gamma$ for the original Abergel and Fal'ko case of
$\mu=0$ (blue dashed curve), thus reproducing their results, and 
our extension to finite $\mu$, specifically $\mu=0.2\gamma$
(solid red curve) and $\mu=1.2\gamma$ (dash-dotted blue). In the two
last
 cases, the two delta functions at
$\Omega=0$ and $\Omega=\gamma$ are shown 
as vertical arrows.
Their weight is given by $a(\mu)/2$ and $b(\mu)$ of Eqs.~(\ref{eq:amu})
and (\ref{eq:bmu}), respectively. Note, that only half of the
delta function at $\Omega=0$ is to be assoicated with the optical
spectral weight for $\Omega\ge 0$. The arrows
in Fig.~\ref{fig1}a are shown schematically
to represent their relative weight.
The quantities $a(\mu)/2$ and $b(\mu)$
are shown in  Fig.~\ref{fig2}  as the long-dashed
blue and short-dashed red curves, respectively, as a function of chemical
potential $\mu$ normalized to the plane-to-plane hopping $\gamma$.
Also, shown for comparison is the case of the uncoupled graphene 
bilayer  with $\gamma=0$ but $\mu$ finite (solid black
curve). We see that the amount of spectral weight in the Drude delta
function centered at $\Omega=0$ is less than it is for pure graphene,
except in the limit of $\mu\to 0$, when they are equal.

Besides the two delta functions, the finite $\mu$ result
has regions of optical spectral weight lost when compared with the
$\mu=0$ case. In the solid red curve of Fig.~\ref{fig1} (top frame), 
the region from 0 to $2\mu$ is completely depleted
while a second region of partial depletion is seen 
above $\Omega=\gamma$. The optical spectral weight lost below
$2\mu$, relative to the $\mu=0$ finite gamma case, can
easily be computed from Eq.~(\ref{eq:falkomu}) and is 
$c(\mu)\equiv\mu+(\gamma/2)\ln[(2\mu+\gamma)/\gamma]$,
which is shown as the dash-dotted green curve in 
Fig.~\ref{fig2}. For finite $\mu$ ($\mu<\gamma$), $c(\mu)>a(\mu)/2$
so that part of the lost spectral weight has been transferred to the second
delta function at 
$\Omega=\gamma$. The existence
of the two regions of depletion just described is encoded in the theta
functions of Eq.~(\ref{eq:falkomu}) and is easily understood in terms of
the energy dispersion curve diagrams displayed in the middle and
bottom frames of Fig.~\ref{fig1}, where
we show possible optical transitions. We start with the middle frame
where the chemical potential, shown as the horizontal
dotted line, is set at $\mu=0$. The transitions are vertical and all
connect a filled valence band to empty conduction band states. From
left to right, the arrows depict typical transitions for  
the four possible ways of connecting the four bands: from
$-\epsilon_2\to+\epsilon_1$  and $-\epsilon_1\to+\epsilon_2$ , which are
both restricted to photon energies $\Omega>\gamma$, and 
$-\epsilon_1\to+\epsilon_1$  and $-\epsilon_2\to+\epsilon_2$, with
the latter one
restricted to $\Omega>2\gamma$.
The onsets of transitions at
$\Omega=\gamma$ and $2\gamma$ are clearly seen in the dashed
blue curve of the top frame of Fig.~\ref{fig1}.

This is to be contrasted with the case of $\mu\ne 0$ shown in the
lower frames. Starting again from the left, the first arrow
describes transitions from $-\epsilon_2\to+\epsilon_1$ as before
but its color has been changed to blue to indicate that now there is
a restriction that only transitions with $\Omega>\gamma+2\mu$
are allowed. The next type of transition remains unchanged for
$\mu<\gamma$, as is the case shown. However, new transitions shown as the
red arrow from $+\epsilon_1\to+\epsilon_2$, can now occur, which were
not possible for $\mu=0$. (The green shading represents
the newly filled states due to finite $\mu$ which were formerly
part of the empty conduction band.) Intraband transitions from
$+\epsilon_1\to+\epsilon_1$  are also possible and correspond to the
delta function centered at $\Omega=0$ which broadens to a Drude
form when any kind of scattering is included. These transitions are
depicted by a short red arrow with vanishing length for $\Omega\to 0$.
The next set from
$-\epsilon_1\to+\epsilon_1$ are restricted to $\Omega>2\mu$ while the
last, $-\epsilon_2\to+\epsilon_2$, remain unaltered. The spectral
weight lost between $0$ and $2\mu$ and between
$\gamma$ and $\gamma+2\mu$ is
balanced
by the appearance of the two delta function contributions
originating from the two new transitions $+\epsilon_1\to+\epsilon_2$ 
and $+\epsilon_1\to+\epsilon_1$ (Drude). The first set of these transitions
leads to a delta function at $\Omega=\gamma$ and this is traced to the
perfect nesting of the bilayer dispersion curves for $\epsilon_1({\bf k})$
and $\epsilon_2({\bf k})$ which differ only by a constant displacement
of $\gamma$. Should the electronic dispersions stop being nested due
to, for example, higher neighbor hoppings, or as we will see,
due to the opening of the semiconducting gap, the delta function will
broaden into a ``band'', the width of which is related to the mismatch
of the dispersions from perfect nesting. 
In particular, as we have already stated, the hopping from A1 to B2
can be significant in size and will lead to changes in the band structure
which are not cylindrically symmetric in the $(k_x,k_y)$-plane\cite{McCann2}.
This fact complicates the mathematics and goes beyond the present
work. However, the trigonal warping which arises can lead to subtle effects
such as described by Mikitik and Sharlai\cite{Mikitik} in relation to
the Dirac points. 

\begin{figure}
\caption{(Color online) Shown to the right.
Upper frame: Frequency dependent conductivity
$\sigma(\Omega)$ of the bilayer normalized to twice that of a  single
  graphene sheet $\sigma_0$ versus $\Omega/\gamma$ for three
values of the chemical potential, as indicated, and $\Delta=0$. Middle frame:
Dispersion of $\epsilon_1$ (solid line)
and $\epsilon_2$ (dashed) of the bilayer near the $K$ point
 with the bands split by $\gamma$, showing the
four types of transitions possible when $\mu=0$.
Bottom frames: Transitions possible in the band structure for finite
chemical potential of $\mu=0.2\gamma$ (lefthand side) and $1.2\gamma$
(righthand side). See text for discussion.
}
\label{fig1}
\end{figure}
\begin{figure}[ht]
\begin{picture}(250,200)
\leavevmode\centering\includegraphics{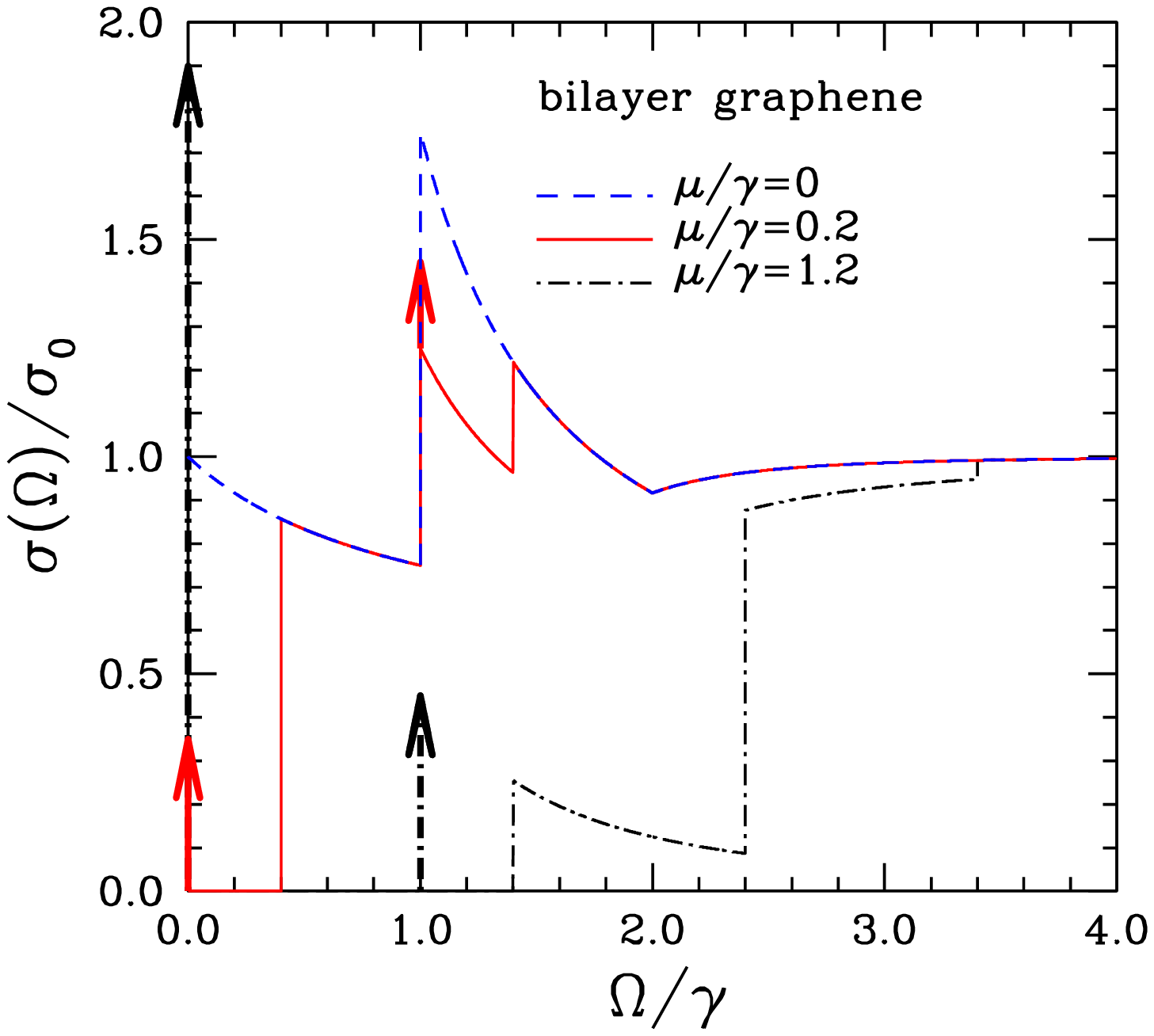}
\end{picture}
\begin{picture}(250,200)
\leavevmode\centering\includegraphics{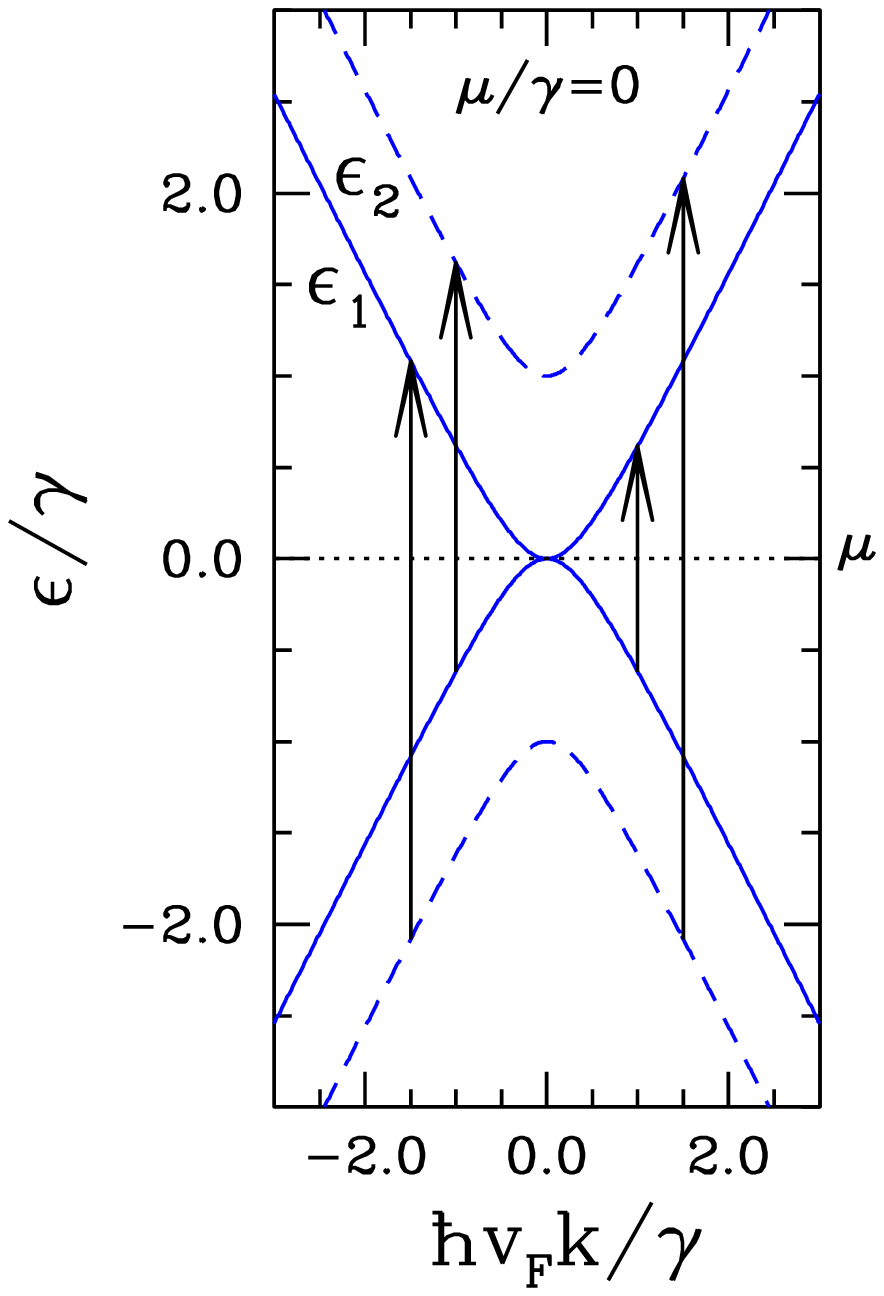}
\end{picture}
\begin{picture}(250,200)
\leavevmode\centering\includegraphics{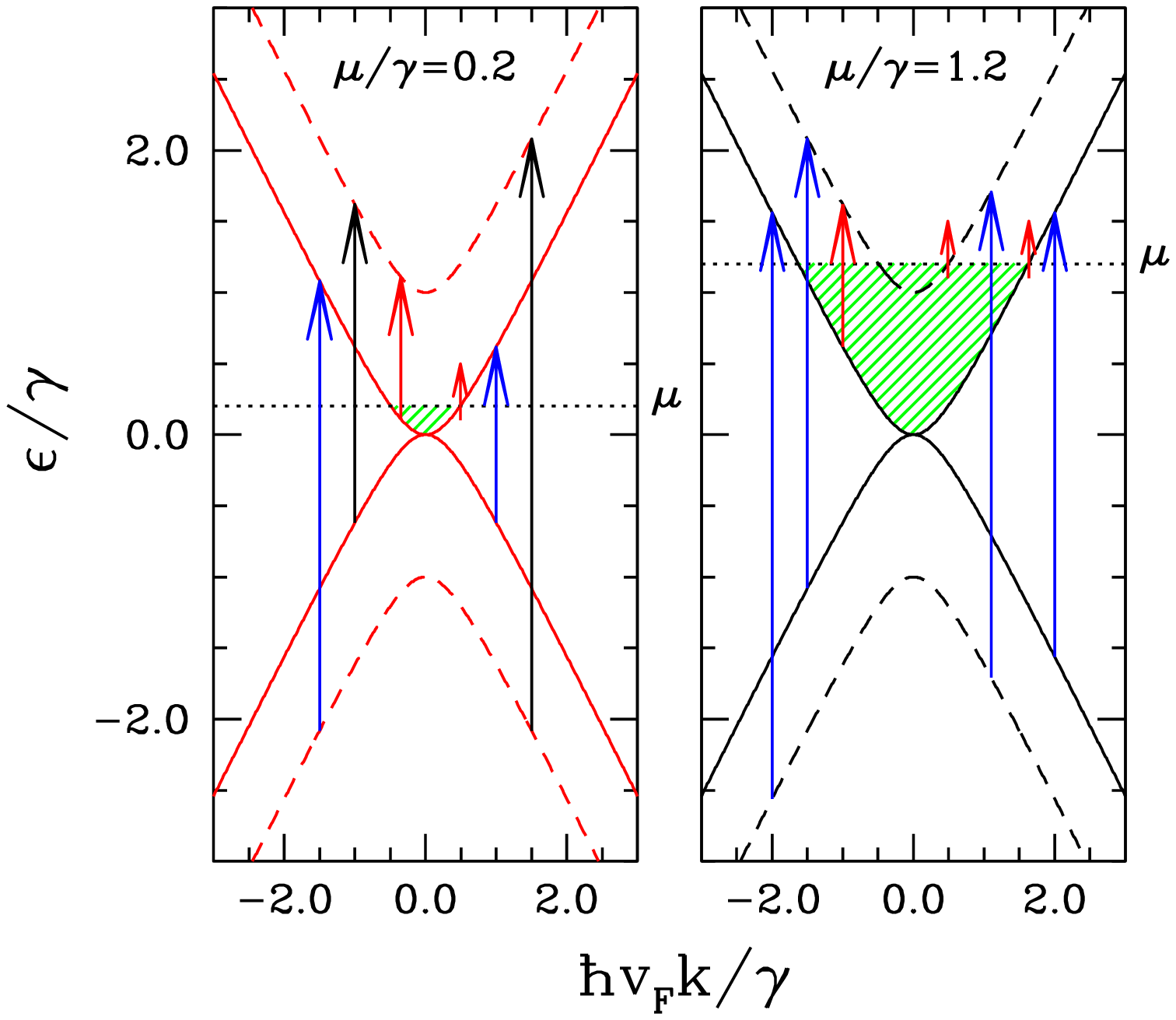}
\end{picture}
\vskip -90pt
\end{figure}

\begin{figure}[ht]
\begin{picture}(250,200)
\leavevmode\centering\includegraphics{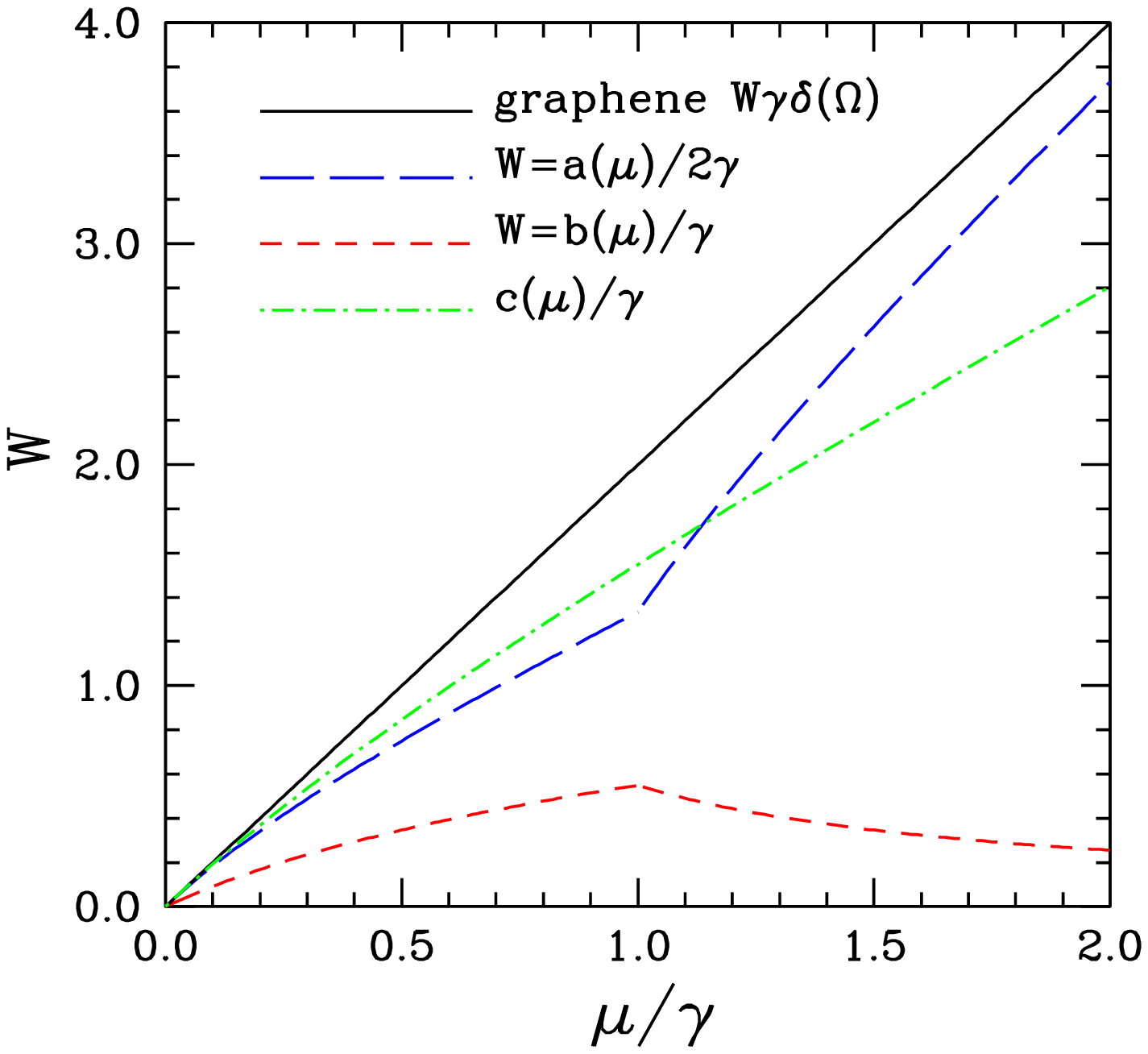}
\end{picture}
\caption{(Color online) The evolution of the positive frequency
spectral weight $W$, found under each of the
delta functions in Eq.~(\ref{eq:falkomu}), as function of chemical potential.
The case of an uncoupled graphene bilayer 
is given for comparison and the quantity $c(\mu)$
represents the spectral weight missing in the conductivity for
$0<\Omega<2\mu$ relative to the $\mu=0$ case.
}
\label{fig2}
\end{figure}

If $\mu> \gamma$, then further new transitions 
become possible. We now show an example for the case where
$\mu=1.2\gamma$ so that the $+\epsilon_2$ band is now
partially occupied. The possible transitions are shown in the bottom
righthand frame of Fig.~\ref{fig1}. 
For the first time all the terms in Eq.~(\ref{eq:falkomu})
become activated. From the diagram $-\epsilon_2\to+\epsilon_1$ transitions
are restricted to $\Omega\ge 2\mu+\gamma$ (first term of second line in Eq.~(\ref{eq:falkomu})) and the restriction on
$-\epsilon_1\to+\epsilon_2$ is 
$\Omega\ge 2(\mu-\gamma)+\gamma=2\mu-\gamma$
(second term of second line in Eq.~(\ref{eq:falkomu})).
 Both the
$-\epsilon_1\to+\epsilon_1$ and $-\epsilon_2\to+\epsilon_2$ arise
only for $\Omega\ge 2\mu$ (first line of Eq.~(\ref{eq:falkomu}))
and new intraband transitions $+\epsilon_2\to+\epsilon_2$ add
the second term in $a(\mu)$. The $b(\mu)$ is depleted through 
an additional negative contribution for $\mu>\gamma$, because the
$+\epsilon_1\to+\epsilon_2$  nested transitions are partially
blocked by the filled states in $+\epsilon_2$ near $k=0$. 
These facts manifest themselves in the top frame of 
Fig.~\ref{fig1} as a large peak centered at $\Omega=0$ and
another at $\Omega=\gamma$
from $a(\mu)$ and $b(\mu)$, respectively. All other transitions
are completely 
suppressed below $\Omega=2\mu=2.4\gamma$ except for a small contribution
from $-\epsilon_1\to+\epsilon_2$ clearly seen between 
$2\mu-\gamma=1.4\gamma$ and $2.4\gamma$.
In addition, the added possible transitions
 above $2\mu+\gamma=3.4\gamma$ are seen to provide a small additional
jump in $\sigma(\Omega)$ at this energy.

\begin{figure}[ht]
\begin{picture}(250,200)
\leavevmode\centering\includegraphics{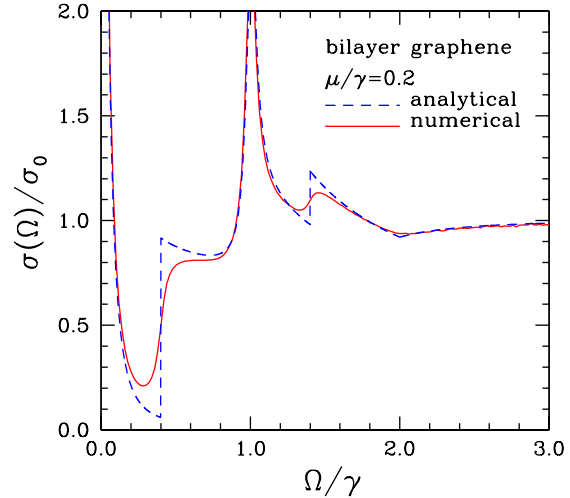}
\end{picture}
\begin{picture}(250,200)
\leavevmode\centering\includegraphics{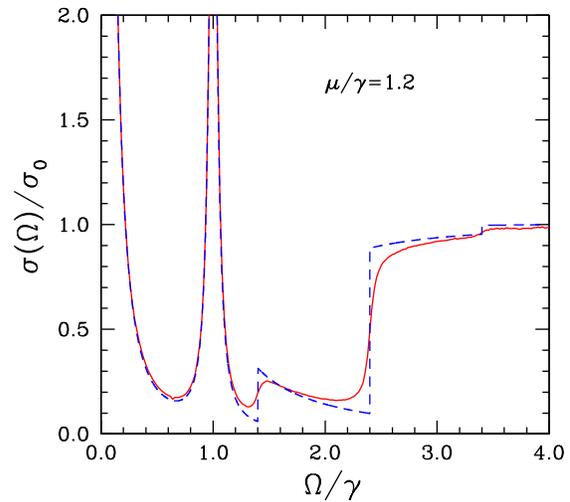}
\end{picture}
\caption{(Color online) Comparison between the full numerical
evaluation of the conductivity [Eq.~(\ref{eq:cond})] and the
analytical formula of Eq.~(\ref{eq:falkomu}). For
Eq.~(\ref{eq:falkomu}) the delta functions are broadened into
Lorentzians with a scattering rate to match the effective
scattering rate in the numerical work. Two different regimes of
$\mu$ are shown: $\mu=0.2\gamma$ (upper frame) and $1.2\gamma$
(lower frame).
}
\label{fig3}
\end{figure}

In Fig.~\ref{fig3}, we compare the
results obtained on the basis of Eq.~(\ref{eq:falkomu}) with the
numerical evaluation of the complete formula Eq.~(\ref{eq:cond}). For
this evaluation a finite value for the electron scattering rate
$1/\tau_{imp}=2\eta$ is
needed. We took it to be a constant in energy and equal to
$0.04\gamma$.
This is the simplest of models and
 is sufficient for our purpose here. In general,
the electron self-energy is a frequency-dependent complex function
with non-vanishing real part. For impurity scattering, the detailed
energy dependence of these functions will depend on 
the scattering potential and will be different for strong (unitary)
and weak (Born) scattering as discussed at length in a recent
preprint\cite{Nilssonnew}. The energy dependence of the scattering
rate
can lead to interesting effects as discussed by Gusynin et
al.\cite{Gusynin3}
for the case of the microwave conductivity where a cusp-like behavior
is predicted to arise insead of the Lorentzian-like dependence
of the usual Drude form. Here, it is sufficient to use a constant
scattering rate in which case the real part of the self-energy,  which
is Kramers-Kronig related, is zero. Returning to the comparison in
Fig.~\ref{fig3}, we note that for the analytic formula
(\ref{eq:falkomu})
we broadened out the delta functions into Lorentzians
using $\delta(x)=(\Gamma/\pi)/(x^2+\Gamma^2)$,
 where $\Gamma=1/\tau_{imp}=0.04\gamma$ to match the full
numerical work. The agreement between the numerical calculation
(solid red curve) and the results of formula (\ref{eq:falkomu})
(dashed blue curve) is excellent. The differences arise solely
because no scattering was included in the dashed
blue curve
beyond broadening the two delta function contributions at $\Omega=0$
and $\Omega=\gamma$ for ease of comparison. 
We have verified that making $1/\tau_{imp}$ smaller brings the two curves
closer, as it must. Note, that concurrent numerical 
work to ours\cite{Nilssonnew}
shows the case of $2\mu=\gamma$
with varying
impurity scattering.
Their results agree with ours.
Similar good agreement between our analytic results based on
Eq.~(\ref{eq:falkomu})
and the full numerical evaluation of Eq.~(\ref{eq:cond}) is seen in
the
lower frame of Fig.~\ref{fig3} for $\mu=1.2\gamma$.
Indeed, we see that broadening the delta functions of the analytical formula,
as we have done, is essential to capture the result that the optical
absorption below $2\mu-\gamma$, in this case, is not zero but actually
finite everywhere. This makes Eq.~(\ref{eq:falkomu}) a useful formula
for experimentalists.

\begin{figure}[ht]
\begin{picture}(250,200)
\leavevmode\centering\includegraphics{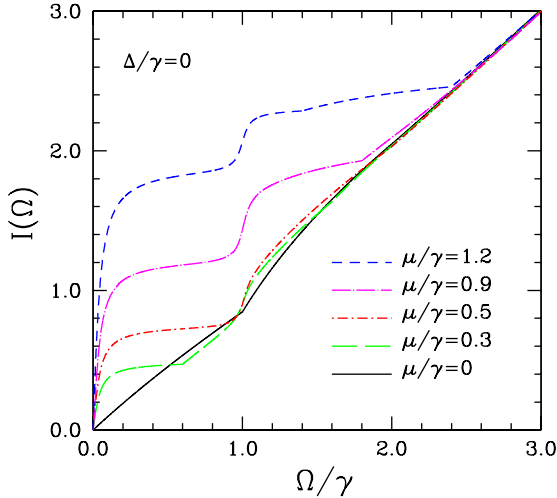}
\end{picture}
\caption{(Color online) Partial optical sum $I(\Omega)$ in units
of $\gamma$ versus
$\Omega/\gamma$ for various values of chemical potential, as indicated
in the figure.
}
\label{fig4}
\end{figure}

The issue of optical spectral weight redistribution with changes in
chemical potential can be addressed in a more global fashion
than we have done so far by introducing the partial optical sum:
\begin{equation}
I(\Omega)=\int_{0^+}^\Omega\frac{\sigma(\omega)}{\sigma_0}d\omega
\label{eq:I}
\end{equation}
defined as the area under the conductivity up to energy $\Omega$.
This is shown in Fig.~\ref{fig4} for five values of the chemical
potential: $\mu/\gamma=0$ (solid black), 0.3 (long-dashed green),
0.5 (short-dashed-dotted red), 
0.9 (long-dash-dotted pink) and 1.2 (short-dashed blue).
In all cases by $\Omega/\gamma=3$, the highest frequency shown,
the integrated spectral weight has returned to its $\mu=0$
value (solid black curve) and the introduction of a finite charge
carrier
imbalance has not changed the partial optical sum up to that
energy, 
although it has significantly changed its distribution in energy in the range
$0<\Omega<3\gamma$. Note in particular the sharp
rise out of $\Omega=0$ exhibited by all curves except for $\mu=0$.
  This reflects the presence of the delta function
contribution at $\Omega=0$ which increases with increasing $\mu$.
The curves start to flatten when most of the spectral weight of the Drude
contribution is integrated and consequently, this plateau occurs at
about the same value as $a(\mu)$ plotted in Fig.~\ref{fig2}.
A second steep rise is also seen at $\Omega=\gamma$ due to the second
delta function in Eq.~(\ref{eq:falkomu}). For the long-dashed green
curve, we note the abrupt change in slope at $\Omega=2\mu=0.6\gamma$
which has moved to $\gamma$ in the short-dash-dotted red curve, to $1.8\gamma$
in the long-dash-dotted pink curve and to $2.4\gamma$ in the short-dashed 
blue curve. These
all reflect the $2\mu$ cutoff. Note also in this last case, the small
kink
at $2\mu-\gamma$ which reflects the onset of the
$-\epsilon_1\to+\epsilon_2$
optical transitions.

\section{Results with anisotropy gap}

\begin{figure}[ht]
\begin{picture}(250,200)
\leavevmode\centering\includegraphics{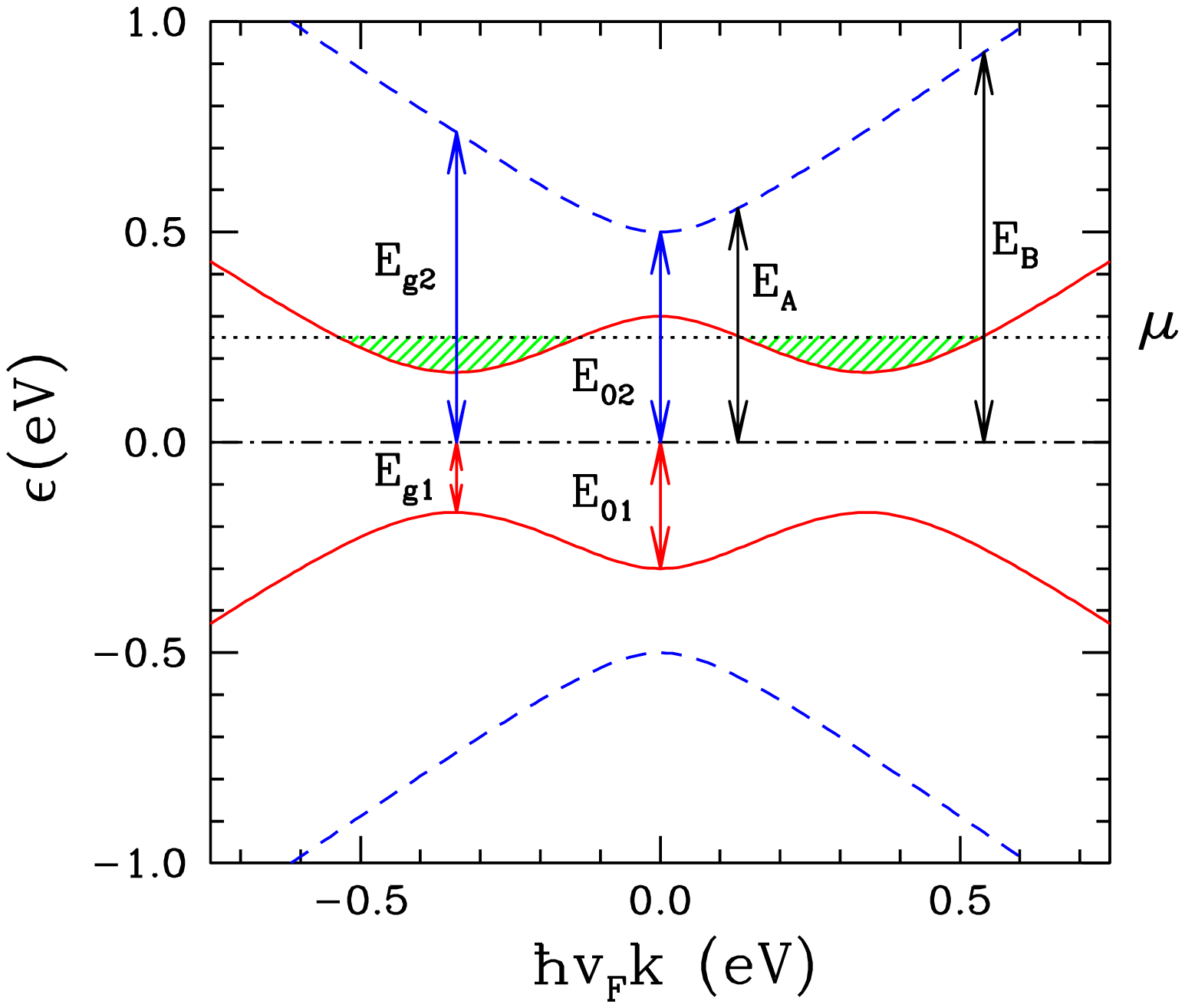}
\end{picture}
\begin{picture}(250,200)
\leavevmode\centering\includegraphics{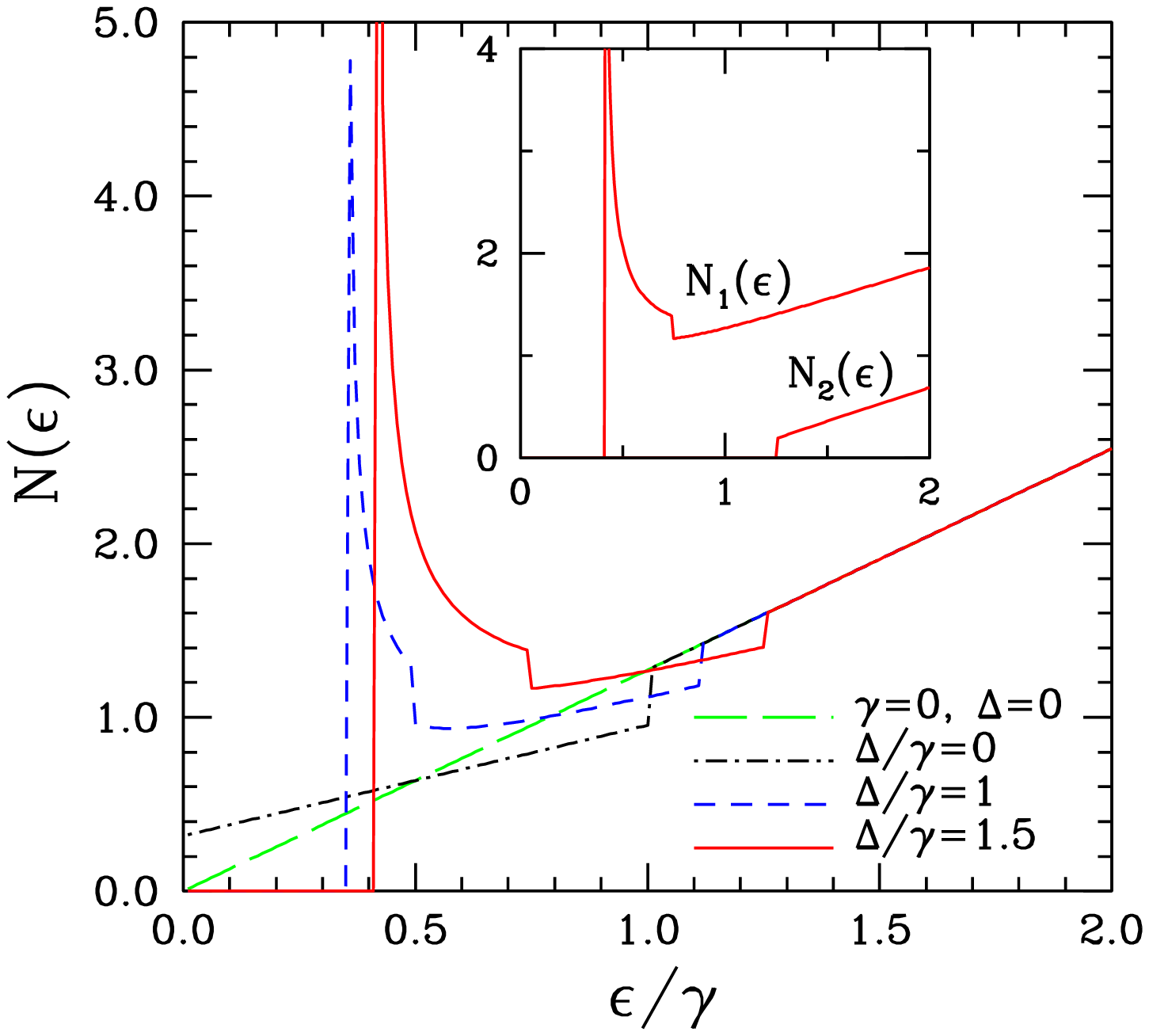}
\end{picture}
\caption{(Color online) Top frame: Band structure around the
Fermi level in the presence
of the asymmetry gap $\Delta$ for realistic values
of $\Delta=0.6$ eV and $\gamma=0.4$ eV. A finite chemical potential of
$\mu=0.25$ eV is shown. Various important energies
are indicated in the figure and displayed in Table~\ref{table1}.
Bottom frame: The density of states $N(\epsilon)$, in units
of $\gamma/\hbar^2v_F^2$,
for the bilayer for several
values of $\Delta$ as indicated in the figure, in comparison
with the case of $\gamma=\Delta=0$ corresponding to the uncoupled graphene
bilayer. The inset shows the partial density of states $N_1(\epsilon)$
and $N_2(\epsilon)$ for the two
separate bands $\epsilon_1$ and $\epsilon_2$, respectively, for the
case
of $\Delta/\gamma=1.5$.
}
\label{fig5}
\end{figure}

For the finite gap case, we begin with a discussion of the
important energy scales involved. In the top frame of Fig.~\ref{fig5},
we show the dispersion curves for our two bands: $\epsilon_1$ (solid red curves) and $\epsilon_2$  (dashed blue
curves). The first band shows a 
mexican hat structure. For positive energies, there is a local maximum
at zero momentum $\phi=\hbar v_F k= 0$ with $\epsilon_1=\Delta/2\equiv E_{01}$.
 There
are also two minimum at finite 
$\phi=\pm(\Delta/2)\sqrt{(\Delta^2+2\gamma^2)/(\Delta^2+\gamma^2)}\equiv\hbar v_Fk_0$ 
with energy $\Delta\gamma/(2\sqrt{\Delta^2+\gamma^2})\equiv E_{g1}$.
The lowest energy for which optical transitions are possible in the semiconducting 
case is $2E_{g1}$,
when there is no charge carrier imbalance which would introduce a finite
$\mu$. For understanding optical transitions, other energies are also
significant and these are indicated in Fig.~\ref{fig5}. At $k=0$,
$\epsilon_2(k=0)=\sqrt{\gamma^2+(\Delta/2)^2}\equiv E_{02}$ and at the momentum
of the minimum in band 1, the energy in band 2 is 
$\epsilon_2(k_0)
=\sqrt{(\Delta^4+\gamma^4+9\Delta^2\gamma^2/4)/(\gamma^2+\Delta^2)}
\equiv E_{g2}$.
In addition, when a finite value of chemical potential is considered with
$\mu>\Delta\gamma/(2\sqrt{\Delta^2+\gamma^2})$ as shown, two other
energies are important, namely the energy $\epsilon_2$ for momenta
at which $\mu$ crosses the $\epsilon_1$ dispersion curve.
These are $\sqrt{\gamma^2+\Delta^2+\mu^2+2L}\equiv E_B$ and if $\mu<\Delta/2$,
$\sqrt{\gamma^2+\Delta^2+\mu^2-2L}\equiv E_A$, where 
$L=\sqrt{\mu^2(\Delta^2+\gamma^2)-(\gamma\Delta/2)^2}$. These quantities
determine the onset of various processes as we will describe below
and are entered in Table~\ref{table1}.
Before doing so, it is important to understand how the electronic density of
states $N(\epsilon)$ varies with $\epsilon$ since its value at the initial and
the final value of the 
energy for a given optical transition provides an important weighting
factor for such processes.

\begin{table*}
\caption{Energies involved in optical transitions (see top frame of 
Fig.~\ref{fig5}) for the case $\gamma=0.4$ eV, $\Delta=0.6$ eV
and two values of chemical potential $\mu=0.25$ and 0.35 eV.}
\begin{ruledtabular}
\begin{tabular}{cccc} 
Energy& Formula& value in eV & value in eV\\
& & for $\mu=0.25$ eV& for $\mu=0.35$ eV\\
 \hline 
$E_{g1}$&$\gamma\Delta/(2\sqrt{\gamma^2+\Delta^2})$&0.166&0.166\\
$E_{g2}$&$\sqrt{(\Delta^4+\gamma^4+9\Delta^2\gamma^2/4)/(\gamma^2+\Delta^2)}$
&0.74&0.74\\
$E_{01}$&$\Delta/2$&0.3&0.3\\
$E_{02}$&$\sqrt{\gamma^2+(\Delta/2)^2}$&0.5&0.5\\
$E_{A}$&$\sqrt{\gamma^2+\Delta^2+\mu^2-2L}$&0.56& -\\
$E_{B}$&$\sqrt{\gamma^2+\Delta^2+\mu^2+2L}$&0.92&1.042\\
&where $L=\sqrt{\mu^2(\gamma^2+\Delta^2)-(\gamma\Delta/2)^2}$&&\\
\end{tabular}
\label{table1}
\end{ruledtabular}
\end{table*}

In the general case, one can obtain an analytic algebraic expression
for total double-spin
$N(\epsilon)$ in terms of the partial density of states provided
by band 1 and band 2, $N_1$ and $N_2$, respectively,
\begin{equation}
N(\epsilon)=N_1(\epsilon)+N_2(\epsilon),
\label{eq:dos}
\end{equation}
where
\begin{eqnarray}
N_1(\epsilon)&=&
[N_1^+(\epsilon)
-N_1^-(\epsilon)\Theta(E_{01}-\epsilon)]
\Theta(\epsilon-E_{g1}),\\
N_2(\epsilon)&=& N_2^-(\epsilon)\Theta(\epsilon-E_{02}),
\label{eq:dos12}
\end{eqnarray}
with 
\begin{widetext}
\begin{eqnarray}
N_\alpha^\pm(\epsilon) &=& \frac{2}{\pi\hbar^2 v_F^2}\frac{\epsilon}{1+
  (-1)^\alpha(\gamma^2+\Delta^2)/C},\nonumber\\
C&=& \sqrt{\gamma^4+(\gamma^2+\Delta^2)[4\epsilon^2+\Delta^2\pm
2\sqrt{4\epsilon^2(\Delta^2+\gamma^2)-\gamma^2\Delta^2}]}.
\label{eq:dospm}
\end{eqnarray}
\end{widetext}
The two terms for $N_1$
simply reflect the two pieces of the energy dispersion for $\epsilon_1$:
the large $k$ piece and the small $k$ piece associated with the
mexican hat. The density of states normalized by $\gamma/\hbar^2 v_F^2$ 
is shown in the bottom frame of
 Fig.~\ref{fig5} for several
values of $\Delta$. The long-dashed green curve is for the uncoupled graphene
bilayer with $\gamma=\Delta=0$
 and is included for comparison. 
 We note that beyond
$\epsilon/\gamma \sim 1.2$
 for the parameters shown here, the various curves come close together
on the scale of the figure. The inset show the density of states associated
with each band separately for the case of $\Delta=1.5\gamma$. 
The $\epsilon_1$ dispersion gives rise to the square root
singularity, followed by a shoulder, that is seen in the total
density of states.
Indeed the singularity in both the solid and short-dashed curves of the main
frame at $\epsilon=E_{g1}$  and the shoulder at $\Delta/2$,
have their origins in the mexican hat structure of the dispersion curve.
The shoulder comes from the top of the hat and is a van Hove singularity
associated with the dispersion flattening at this point.
The square root singularity can be derived for $\epsilon$ near the hat
minimum $\epsilon\sim E_{g1}$ and we find, as others\cite{Nilssonnew}
have done,
\begin{equation}
N(\epsilon)=\frac{k_0}{4\pi\hbar}\sqrt{\frac{2m^*}{\epsilon-E_{g1}}}.
\label{eq:dossing}
\end{equation}
The factor $k_0=(\Delta/2\hbar v_F)\sqrt{(\Delta^2+2\gamma^2)/(\Delta^2+\gamma^2)}$, which is the momentum associated with
the rim of the mexican hat,
 is related to the degeneracy of energies around the circular
minimum of the mexican hat (fixed magnitude, varying angle of momentum)
while the square root is associated with the 
one-dimensional variation in energy up and down the rim with varying
magnitude of
$k$ (fixed angle), 
making this equivalent to what is expected for a one-dimensional
density of states. The effective mass is $m^*=\frac{\gamma}{2\Delta v^2_F}
\frac{[\Delta^2+\gamma^2]^{3/2}}{\Delta^2+2\gamma^2}$
with energy $\epsilon = E_{g1}+\frac{\hbar^2}{2m^*}(k-k_0)^2$. 
For comparing with Eq.~(\ref{eq:dos}), it is convenient to rewrite
Eq.~(\ref{eq:dossing}) in the form:
\begin{equation}
N(\epsilon)=\frac{1}{8\pi\hbar^2v_F^2}\sqrt{\frac{\gamma\Delta\sqrt{\Delta^2+\gamma^2}}{\epsilon-E_{g1}}}.
\end{equation}
This singularity
plays an important role in optics because it leads to peaks in $\sigma(\Omega)$
when the energy of either or both the initial and final states involved falls at
$E_{g1}$. This fact, plus the energy scales identified in Fig.~\ref{fig5}
(top frame) allows one to understand the qualitative features of our numerical
results for the conductivity.

\begin{figure}[ht]
\begin{picture}(250,200)
\leavevmode\centering\includegraphics{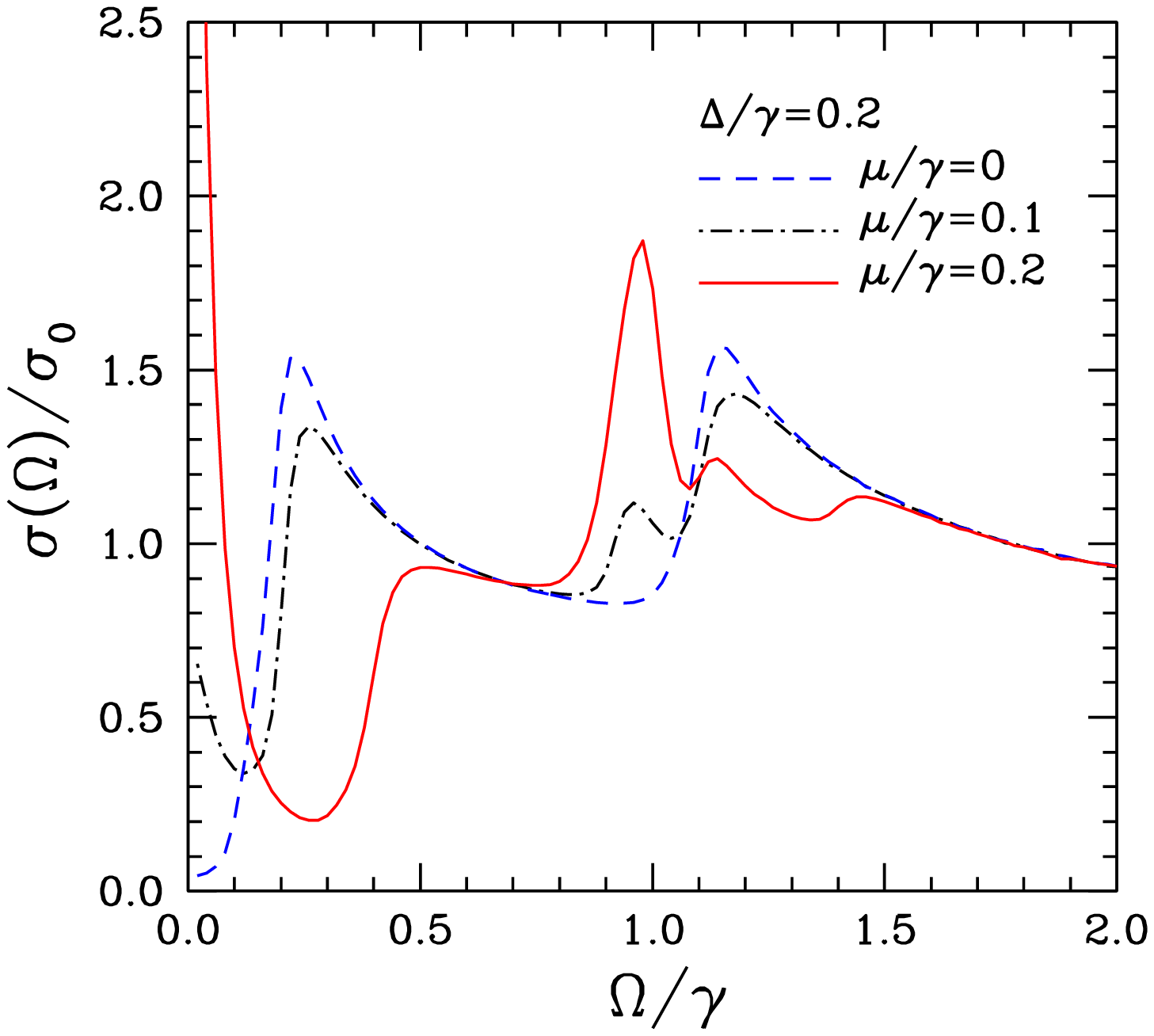}
\end{picture}
\begin{picture}(250,200)
\leavevmode\centering\includegraphics{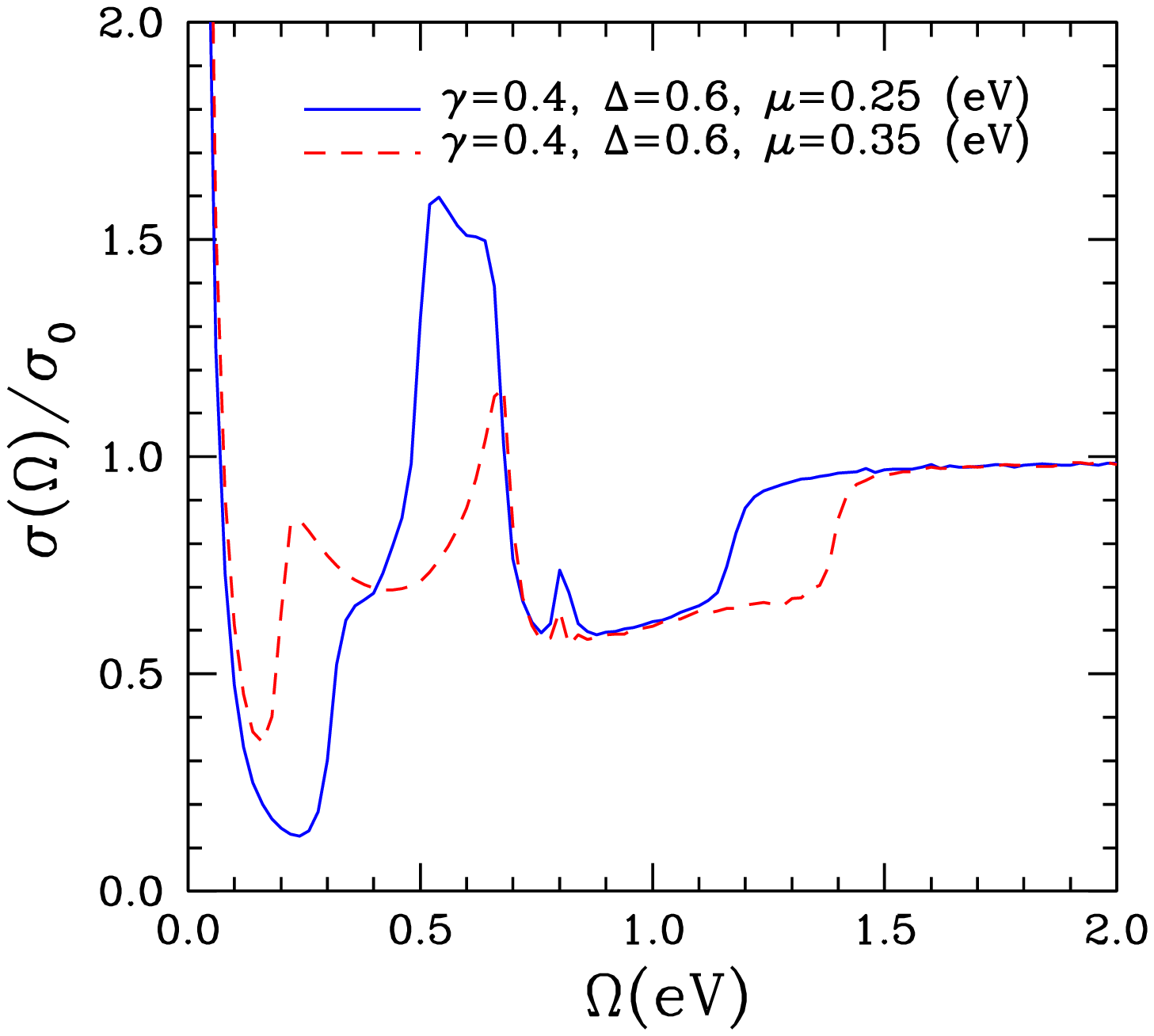}
\end{picture}
\caption{(Color online) Optical conductivity in the presence of
an asymmetry gap $\Delta$. The upper frame shows the results for
varying $\mu$ with fixed $\Delta=0.2\gamma$. The bottom frame
shows the result for realistic values of the parameters, i.e.,
 $\Delta=0.6$
eV and $\gamma=0.4$ eV. Here two curves are shown, one for $\mu=0.25$
eV where the chemical potential lies above the gap in the
band structure
but below the hat maximum, and $\mu=0.35$ eV, where it lies above the 
hat maximum. 
}
\label{fig6}
\end{figure}

Referring to Fig.~\ref{fig6}, we start with the upper frame,
where $\Delta=0.2\gamma$. Three values of the chemical potential
$\mu$ are considered. The dashed blue
 curve, which is for reference, has $\mu=0$,
the dash-dotted black is for $\mu=0.1\gamma$ and the solid red
 is for $\mu=0.2\gamma$. We begin
with the dashed blue curve which is the only case which exhibits a true
semiconducting gap with no absorption up to photon energy 
$\Omega=2E_{g1}=0.196\gamma$.
We first note the existence  of 
a small tail below this energy and also note that the
main rise is smeared. This is
 because we have used  a finite impurity scattering
rate of $1/\tau_{imp}=0.04\gamma$ in all of 
our numerical work. It is clear that most
of the states lost below the gap ($2E_{g1}$) are to be found piled
up in a region of order a few $E_{g1}$ above it. Later we will
examine in detail this optical spectral weight redistribution. 
This part
of the curve for $\sigma(\Omega)$ is due to the $-\epsilon_1\to+\epsilon_1$
transitions and their
onset involves the singular density of states at 
$E_{g1}$ for both initial and final states. The second peak 
in the dashed blue
curve is traced to transitions $-\epsilon_2\to+\epsilon_1$
and $-\epsilon_1\to+\epsilon_2$ (degenerate in energies)
and these involve a single singular 
$N(\epsilon)$ either in the initial or final state but not for both,
and the onset is at $E_{02}+E_{01}=
\sqrt{\gamma^2+(\Delta/2)^2}+\Delta/2\simeq 1.1\gamma$.
It is also clear from Fig.~\ref{fig5} (top frame) that this onset remains 
even for finite $\mu$ because the $-\epsilon_1\to+\epsilon_2$ 
transition is never blocked by a finite $\mu$ value provided $\mu$ falls below 
the minimum in $\epsilon_2$ which is the case considered here.

Next we consider the dash-dotted black curve for $\mu=0.1\gamma$. In this case,
the $-\epsilon_1\to+\epsilon_1$ transitions are no longer possible for energies
less than or equal to $2\mu=0.2\gamma$ because these states are occupied
and cannot be used as final states in optical transitions. The energy
$2\mu=0.2\gamma$ 
is very close to the onset for the dashed blue curve with $\mu=0$ which
is at $0.196\gamma$ and,  as a result,
both dashed and dash-dotted curves are close to each other in this
energy region.
Nevertheless it is important to understand that while
the dashed blue curve 
is in principle encoded with the information on the details 
of the mexican hat topology, the dash-dotted black curve is less so
 as it involves,
in addition, a sharp cut off at $0.2\gamma$,
which obscures some of these details. Finally we note the transfer of optical
spectral weight to a Drude peak centered at $\Omega=0$ due to 
$+\epsilon_1\to+\epsilon_1$ intraband
transitions. This peak does not exist at zero
temperature for $\Delta\ne 0$ and $\mu=0$.

The next interesting feature in the dash-dotted black 
curve is the peak slightly
above $\Omega=0.9\gamma$. 
This peak can be traced to new $+\epsilon_1\to+\epsilon_2$
optical transitions made possible for finite $\mu$. We have
already seen that, for $\Delta=0$, these would fall at
$\Omega=\gamma$ and provide, in the clean limit, a delta function contribution
which can be traced to the perfect nesting of the dispersion curves. But 
for finite $\Delta$, perfect nesting no longer occurs and these
transitions broaden into a ``band'' as well as shift in energy as we now describe.
Returning to Fig.~\ref{fig5} (top frame), we note that because $+\epsilon_1$
states are now occupied in a region about the mexican hat minimum,
 transitions from $+\epsilon_1\to+\epsilon_2$ become possible for photon
energies between $E_A-\mu$
and $E_B-\mu$. The intermediate
energy transition at
$E_{g2}-E_{g1}$
involves the density of states at the bottom of the mexican hat
and this is expected to lead to a peak in $\sigma(\Omega)$ at this
energy. For the parameters of the model, the new ``band'' ranges
in energy between $0.915\gamma$ to $0.934\gamma$, with a peak at $0.93\gamma$.
 In Fig.~\ref{fig6},
 an additional broadening of this narrow ``band''
is included because we have used a finite
$1/\tau_{imp}$. Nevertheless, overall, the new ``band'' 
does not broaden much as a result of scattering
and the changed topology of the energy dispersion curves
$\epsilon_1({\bf k})$ and $\epsilon_2({\bf k})$. While
the dispersion curves $+\epsilon_1$ and $+\epsilon_2$
are no longer simply displaced by a constant amount with respect to each other
and therefore are no longer perfectly nested, the effect is not large.

As we have already discussed, the $-\epsilon_1\to+\epsilon_2$ transitions
remain unaffected by a finite but small value of
$\mu$ and the onset for these
transitions remains unshifted in energy at $E_{01}+E_{02}\simeq 1.1$. 
However, the optical
weight at the onset is depleted 
because the $-\epsilon_2\to+\epsilon_1$ transitions do know about $\mu$.
For $\mu=0.1\gamma$, these transitions are blocked in the energy range
$E_A+\mu=1.115\gamma$ to
$E_B+\mu=1.134\gamma$. Both these energies fall 
very close to the peak energy $1.1\gamma$ and show up in the figure simply
as a slight depletion of the large broad peak above $\Omega=1.1\gamma$ in the
dashed blue curve.

Similar arguments explain the main features seen in the solid red curve
for $\mu=0.2\gamma$. In this case, the cutoff imposed on the 
$-\epsilon_1\to+\epsilon_1$ transitions 
is at $\Omega=0.4\gamma$ ($2\mu$) and the amount
of spectral weight transferred to the Drude centered at $\Omega=0$
has greatly increased as compared with the dash-dotted black curve. The peak
from the new $+\epsilon_1\to+\epsilon_2$ transitions has broadened 
extending from $0.9\gamma$, set by $E_{02}-E_{01}$ (as $\mu$ is now above the
hat maximum) to
$0.998\gamma$, set by $E_B-\mu$.
 Finally, the upper cutoff on the $-\epsilon_2\to+\epsilon_1$ transitions
has moved to $E_B+\mu=1.4$ eV, where a sharp rise in conductivity is seen.

In a field effect device with a semiconducting gap, the value
of $\Delta$ cannot be set independently from the value of the chemical
potential $\mu$ which sets the occupation of the $+\epsilon_1$ band and
$E_{g1}$, the energy of the 
minimum of the mexican hat. Realistic values based on numerical
work presented in Ref.~\cite{Benfatto} are $\gamma=0.4$, $\Delta=0.6$,
and $\mu=0.35$. Results for the conductivity $\sigma(\Omega)/\sigma_0$ versus
$\Omega$ are shown as the dashed red curve in the bottom frame of
Fig.~\ref{fig6}.
We found it illuminating also to present for
comparison the case of $\mu=0.25$ represented by the solid blue curve.
The band structure involved is shown in the top frame of Fig.~\ref{fig5},
 where various energies
of importance for optical transitions are identified. 
Actual numbers for the $E_i$'s are found in Table~\ref{table1}.
All the important 
qualitative features seen in the conductivity curves 
can be understood on the basis
of these energy scales.
We begin with the solid blue curve. If the chemical potential was zero
the $-\epsilon_1\to+\epsilon_1$ transitions would start at $2E_{g1}=0.333$ eV
but for $\mu=0.25$ these are cutoff at $0.5$ eV where we see a sharp rise
in conductivity. On the other hand, the $+\epsilon_1\to+\epsilon_2$ transitions,
made possible through the finite occupation of the $+\epsilon_1$
states at finite chemical potential $\mu$, start at $E_A-\mu=0.31$ eV
where we see the first rise in $\sigma(\Omega)$ after the Drude.
These transitions extend to $E_B-\mu=0.67$ eV where the conductivity
shows a steep drop. Thus, the nested transitions
of the $\Delta=0$ bilayer graphene case are now greatly broadened
and their onset has moved below the onset for
the $-\epsilon_1\to +\epsilon_1$ transitions.
 The small peak at $E_{01}+E_{02}=0.8$ eV is traced to the 
$-\epsilon_2\to+\epsilon_1$ and $-\epsilon_1\to+\epsilon_2$ transitions
which, as we have discussed, are depleted by charging but their onset 
remains unshifted. Finally the rise at $E_B+\mu=1.17$ eV coincides with the
upper cutoff on the $-\epsilon_2\to+\epsilon_1$ transitions.
For $\mu=0.35$ eV (dashed red curve) the $+\epsilon_1\to+\epsilon_2$ 
transitions start at $E_{02}-E_{01}=0.2$ eV and cause the first rise
in conductivity which follows the Drude
peak about $\Omega=0$. They extend to up $E_B-\mu=0.69$
 eV where the conductivity
shows a rapid decrease. 
We stress that the onset of these transitions is set by the difference
$(E_{02}-E_{01})$ (bottom of second band minus
top of mexican hat) and not directly
by the chemical potential which falls above $E_{01}$.
The peak at 0.8 eV is further depleted as compared
with the solid blue curve but is unshifted. The upper cutoff on the 
$-\epsilon_2\to+\epsilon_1$ transitions is now at $E_B+\mu=1.39$ eV.

\begin{figure}[ht]
\begin{picture}(250,200)
\leavevmode\centering\includegraphics{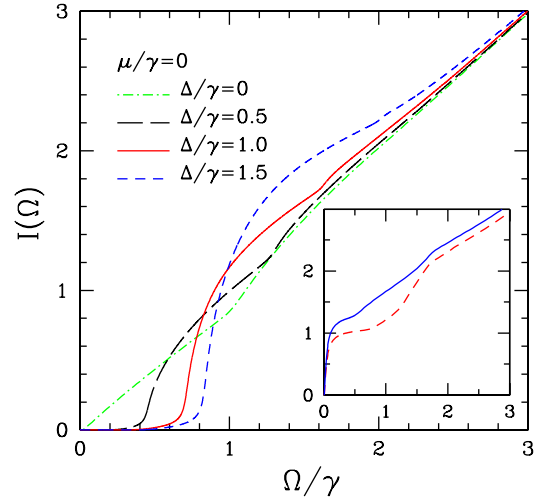}
\end{picture}
\caption{(Color online) The partial optical sum $I(\Omega)$ in
units of $\gamma$
versus $\Omega/\gamma$ for various values of $\Delta$ with $\mu=0$.
Inset: $I(\Omega)$
for the more realistic parameters used for
$\sigma(\Omega)$ in Fig.~\ref{fig6}b, i.e., $\gamma=0.4$ eV, $\Delta=0.6$
eV, and $\mu=0.25$ eV (dashed red curve) and $0.35$ eV (solid blue curve).
}
\label{fig7}
\end{figure}

In Fig.~\ref{fig7}, we show results for the partial optical sum
$I(\Omega)$ in units of $\gamma$ versus $\Omega/\gamma$ which describes the optical weight
redistribution brought about by the opening of the anisotropy gap
and finite chemical potential. Four cases are compared in the main
frame for $\mu=0$. The dash-dotted green
curve is for $\Delta=0$, long dashed black for $\Delta/\gamma=0.5$,
solid red for $\Delta/\gamma=1.0$, and short dashed blue for $\Delta/\gamma=1.5$. 
 We see the
almost complete depletion of optical spectral weight below the
gap $2E_{g1}$ set by the minimum
energy in the mexican hat dispersion curves. The small tails below 
this cutoff are due to the impurity smearing caused by our use of a finite
$1/\tau_{imp}$ in the numerical work. In all cases, the lost
spectral weight is recovered in the region above it. The overshoot beyond
the reference curve (dash-dotted green) is largest for the largest $\Delta$ considered 
and in this case $I(\Omega)$ is still slightly above the $\Delta=0$ reference
even at $\Omega=3$ eV. It is clear, however, that there is a close
balance between weight lost and gained in this interval. 
In the inset of Fig.~\ref{fig7}, we show $I(\Omega)$ in units of
$\gamma$ versus $\Omega/\gamma$ for more realistic
values of $\gamma$, $\Delta$, and $\mu$ associated with the conductivity
curves shown in the bottom
frame of Fig.~\ref{fig6}. For $\mu$ finite, there is
a narrow Drude peak in $\sigma(\Omega)$ with width set by $1/\tau_{imp}$,
which results in a fast rise of $I(\Omega)$ out of $\Omega=0$. This
is followed by further rises with small kinks reflecting the sharp rises and drops
in the associated conductivity curves.

\section{Summary and Discussion}

We have derived an analytic algebraic formula for the zero
temperature, clean limit optical conductivity of a graphene bilayer
valid for any value of charge imbalance characterized by
a general value of the chemical potential $\mu$. In the limit of
$\mu=0$ our formula reduces properly to that of Abergel and Fal'ko\cite{Abergel}.
A finite $\mu$ introduces several important modifications.
First, some of the low energy $-\epsilon_1\to+\epsilon_1$ 
transitions are no longer allowed because the charge imbalance
leads to finite occupation of part of the $+\epsilon_1$ band. Furthermore,
new $+\epsilon_1\to+\epsilon_1$ intraband transitions are now possible
and this leads to a finite Drude peak at $\Omega=0$.
 In addition, new interband transitions between
 $+\epsilon_1\to+\epsilon_2$ are allowed and these lead to a 
second delta function contribution at $\Omega=\gamma$. These transitions
reflect the perfect nesting between bands 1 and 2 which are simply
displaced in energy by a constant amount $\gamma$ in our model
band structure when there is no anisotropy
gap. For more complicated models in which the nesting ceases to be 
perfect, the peak at $\Omega=\gamma$ would broaden into an
absorption band whose
width and structure in energy reflects the mismatch in topology between
bands 1 and 2. In addition, there is a broadening brought about by the elastic
scattering rate $1/\tau_{imp}$. Immediately above the $\Omega=\gamma$ peak 
there is a second depletion region which is due to the blocking of the
$-\epsilon_2\to+\epsilon_1$ transitions between $\gamma$ and $\gamma+2\mu$.
The presence of a finite $\mu$ does not affect the $-\epsilon_1\to+\epsilon_2$ 
transitions which also fall in the same energy interval and so only
partial depletion is involved. It is found that the optical spectral weight
lost in the two depletion regions is completely compensated for by the spectral
weight which resides in the delta function. We have just described the case for 
$\mu<\gamma$. For $\mu>\gamma$, new intraband transitions associated with
the $+\epsilon_2\to+\epsilon_2$ transitions become possible and the transitions
which utilize band 2 as the final state also become blocked.

Note that all the main features of the conductivity curves found here
can be traced directly to the underlying band structure of bilayer 
graphene. 
Correlation effects, which have not been treated here, lead
to self-energy $(\Sigma)$ corrections. The real part of $\Sigma$
renormalizes the single particle energies and the imaginary part
introduces damping. Angle-resolved photoemission spectroscopy (ARPES)
provides a direct measure of the dressed dispersion curves and of
their many body broadening. The ARPES data\cite{Ohta,Bostwick}
 confirms the general shape of
the tight-binding dispersion curves used here with some smaller
modification and by implication, we do not expect large changes
in the frequency dependence of the conductivity described here.
Of course, the scattering rate due to many body interactions will,
in general, be frequency dependent while here we have treated
it as a constant. In bilayer graphene, even impurity scattering
involves an energy-dependent scattering rate and many examples
of how this changes the shape of the optical conductivity can be
found in the recent work by Nilsson et al.\cite{Nilssonnew}
Finally, we note that vertex corrections have not 
been included here as this goes beyond the present work.

When a semiconducting gap is introduced, the band structure becomes
modified. In particular, band 1 acquires a mexican hat structure with the
top of the hat at energy $E_{01}=\Delta/2$ and the minimum
on the rim of the hat is at $E_{g1}=(\Delta\gamma/2)/\sqrt{\Delta^2+\gamma^2}$.
This can be achieved in a graphene bilayer when donor atoms are seeded on its
top surface and the whole is placed in a field effect 
configuration. In such junctions, finite $\Delta$ also implies
finite $\mu>E_{g1}$ and there exists a finite
charge imbalance. Only for the case of finite $\Delta$ and $\mu=0$ is there
a real gap in the system, i.e. no absorption below
$2E_{g1}$: the 
$-\epsilon_1\to+\epsilon_1$ transitions become gapped and the lost optical
spectral weight is found to accumulate in the energy region just above the
gap on an energy scale of order $\Delta$. In any practical case, however,
a finite value of chemical potential accompanies a finite 
$\Delta$ and there is a Drude centered at $\Omega=0$ and a cutoff on the
$-\epsilon_1\to+\epsilon_1$ transition of $2\mu$. The onset of the
$-\epsilon_1\to+\epsilon_2$ (and $-\epsilon_2\to+\epsilon_1$)
moves from $\gamma$ in the case of $\Delta=0$ to
 $E_{01}+E_{02}=\sqrt{(\Delta/2)^2+\gamma^2}+(\Delta/2)$. 
The $-\epsilon_2\to+\epsilon_1$ transitions become gapped in the interval
$E_A+\mu$ to $E_B+\mu$, unless $\mu>\Delta/2$, then the lower limit
changes to $E_{01}+E_{02}$.
In addition, the transitions
from $+\epsilon_1\to+\epsilon_2$, which are possible for finite $\mu$
and which 
in pure unbiased bilayer graphene provide a delta function at $\gamma$,
are broadened into a band from
$E_A-\mu$ to $E_B-\mu$, the width of which depends on the value
of chemical potential. This band of absorption starts at energies 
below $\gamma$,
with onset at the energy of the top of the mexican hat if $\mu$ is
greater than this energy.
While these modifications in the possible
transitions due to finite $\Delta$ and $\mu$ can lead to complicated 
spectral weight shifts in $\sigma(\Omega)$ versus $\Omega$, all changes
can be understood qualitatively from a knowledge of the band structure involved.
Biased graphene bilayers offer a rich pattern of variation of $\sigma(\Omega)$
versus $\Omega$ as the size of the anisotropy gap is varied through changes in 
the voltage of the field effect device. This system is not
only important because of possible practical
applications, but it is also the only system 
known for which the value of the semiconducting gap can be tuned by the
application
of an external voltage.

\begin{acknowledgments}
This work has been supported by NSERC of Canada
and by the Canadian Institute for Advanced Research (CIFAR).
J. P. C. thanks S. Sharapov and L. Benfatto for instructive
discussions. 
\end{acknowledgments}

\end{document}